\documentstyle[12pt]{article}

\begin{document}

\title
{On combining significances. Some trivial examples.}

\author{N.V.Krasnikov and S.I.Bityukov 
  \\
INR RAN, Moscow 117312}

\maketitle
\begin{abstract}

For Poisson distribution $Pois(n, \lambda)$ with 
$\lambda \gg 1$, $n \gg 1$ 
we propose to determine significance as 
$S = \displaystyle \frac{n_{obs}-\lambda}{\sqrt{\lambda}}$. 
The significance $S$ coincides up to sign with often used significance. 
For experiments which measure the same quantities the natural but 
not unique rule for significance combining is 

$S_{comb}(S_1, S_2) = 
\displaystyle \frac{S_1\sigma_1+S_2\sigma_2}{\sqrt{\sigma^2_1+\sigma^2_2}}$,

\noindent
where $\sigma_1$ and $\sigma_2$ are variations. 
We also propose the rule for significances combining for the  case with systematic errors. 
\end{abstract}

\newpage

``Suppose one experiment sees a 3-sigma effect and 
another sees a 4-sigma effect. What is combined 
significance?''~\cite{1}.

In this note we discuss the problem of significances combining on 
the example of Poisson distribution\footnote{See also \cite{2,3}}.  
Namely for Poisson distribution $Pois(n, \lambda)$ with 
$\lambda \gg 1$, $n \gg 1$ 
we propose to determine significance as 
$S = \displaystyle \frac{n_{obs}-\lambda}{\sqrt{\lambda}}$. 
The significance $S$ coincides up to sign with often used significance. 
For experiments which measure the same quantities the natural 
but not unique rule for significance combining is 

$S_{comb}(S_1, S_2) = 
\displaystyle \frac{S_1\sigma_1+S_2\sigma_2}{\sqrt{\sigma^2_1+\sigma^2_2}}$,

\noindent
where $\sigma_1$ and $\sigma_2$ are variations. 
We also propose the rule for significances combining for  systematic errors. 


Suppose the CMS experiment  detects in July 2010  10300 events 
with the expectation 10000 events and in august it detects 9700 events 
with the expectation 10000 events. The probability to detect $n$ events 
is determined by Poisson formula  

\begin{equation}
Pois(n,\lambda) = \displaystyle \frac{\lambda^n}{n!}e^{-\lambda}. 
\label{eq:1} 
\end{equation}

For $\lambda \gg 1$, $n \gg 1$ Poisson distribution is approximated by 
normal distribution with mean $\mu=\lambda$ and variance 
$\sigma = \sqrt{\lambda}$.

For   $\lambda \gg 1$, $n_{obs} \gg 1$ the often used significance 
$S$ is determined by the approximate formula

\begin{equation}
S = \displaystyle \frac{|n_{obs}-\lambda|}{\sqrt{\lambda}}. 
\label{eq:2} 
\end{equation}

For our example we find the july and august CMS significances 

\begin{equation}
S_{july} = \displaystyle \frac{|10300-10000|}{\sqrt{10000}}=3, 
\label{eq:3} 
\end{equation}

\begin{equation}
S_{august} = \displaystyle \frac{|9700-10000|}{\sqrt{10000}}=3, 
\label{eq:4} 
\end{equation}

If we deal with collection of CMS data for july plus august we find 
that data are described by Poisson distribution with 
$\lambda_{july+august}=\lambda_{july}+\lambda_{august}$ and 
$n_{obs,july+august}=n_{obs,july}+n_{obs,august}$.
In accordance with the  formula (2) we find that july + august significance is 

\begin{equation}
S_{july+august} = \displaystyle 
\frac{|n_{obs,july}+n_{obs,august}-\lambda_{july}-\lambda_{august}|}
{\sqrt{\lambda_{july}+\lambda_{august}}}  = 0
\label{eq:5} 
\end{equation}
in perfect agreement with the 
theory expectations. This trivial example illustrates the fact that 
it is impossible to combine only significances, i.e. 

\begin{equation}
S_{july+august} \ne F(S_{july}, S_{august})
\label{eq:6} 
\end{equation}

One of the possible solutions is to determine significance as 

\begin{equation}
S = \displaystyle \frac{n_{obs}-\lambda}{\sqrt{\lambda}}. 
\label{eq:7} 
\end{equation}

The definition (\ref{eq:7}) coincides with often used significance 
definition up to sign. For the case of events excess it is positive 
and for the opposite case  it is negative. 
For the definition (\ref{eq:7}) the proposed rule for significances 
combining is 

\begin{equation}
\displaystyle 
S_{comb}(S_1,S_2) = 
\frac{S_1 \sqrt{\lambda_1} + S_2 \sqrt{\lambda_2}}{\sqrt{\lambda_1+\lambda_2}}
= \frac{S_1 \sigma_1 + S_2 \sigma_2}{\sqrt{\sigma^2_1+\sigma^2_2}}.
\label{eq:8} 
\end{equation}

The rule (\ref{eq:8}) is in fact trivial generalization of Stouffer's 
method~\cite{4} and for Poisson statistics it looks rather natural\footnote{ 
The rule (8)  is the realization of  the fact that the sum of Poisson processes with $\lambda_1$ 
and $\lambda_2$ is the Poisson process with $\lambda=\lambda_1+\lambda_2$.}. 
The generalization of the asymptotic formula (\ref{eq:8}) to the case 
of not big $\lambda_1$ and $\lambda_2$ is straightforward. 


Other lesson from our example and formula (\ref{eq:8}) is that the 
significance combining not necessary leads to increase of the 
significance.

Note that the situation changes if we investigate the case when 
the parameter $\lambda$ in Poisson distribution is not known.
For our example we find that 
$\lambda_{july}=10300,~\lambda_{august}=9700$ and 
$\lambda_{july+august}=20000$. In the assumption that 
$\lambda_{july}=\lambda_{august}$ the analysis of july+august data 
gives $\lambda_{july}=\lambda_{august}=10000$


Other example is the case when CMS detects in july 10300 events 
with the expectation 10000 and in august it detects 10100 events 
with the expectation 10000.

Again we find 

\begin{equation}
S_{july} = \displaystyle \frac{10300-10000}{\sqrt{10000}}=3, 
\label{eq:9} 
\end{equation}

\begin{equation}
S_{august} = \displaystyle \frac{10100-10000}{\sqrt{10000}}=1, 
\label{eq:10} 
\end{equation}

And for CMS data for july+august we find 

\begin{equation}
S_{july+august} = \displaystyle 
\frac{400}{\sqrt{20000}} = \frac{4}{\sqrt{2}} \approx 2.8. 
\label{eq:11} 
\end{equation}

It is not difficult  to take into account systematic effects related with 
nonexact knowledge of the parameter $\lambda$ in 
Poisson distribution (\ref{eq:1}). Namely, suppose theoretical uncertainty in 
the  $\lambda$ parameter calculation is  $\epsilon\lambda$. 
For such uncertainty the generalization of the formula (7) is

\begin{equation}
S = \displaystyle \frac{n_{obs}-\lambda}{\sqrt{\lambda +    (\epsilon\lambda})^2   }. 
\label{eq:12} 
\end{equation}
The generalization of the formula (8) looks

\begin{equation}
\displaystyle 
S_{comb}(S_1,S_2) = 
\frac{S_1 \sqrt{\lambda_1 + (\epsilon\lambda_1)^2} + 
S_2 \sqrt{\lambda_2 + (\epsilon\lambda_2)^2}}{\sqrt{\lambda_1+\lambda_2 +(\epsilon(\lambda_1 + \lambda_2))^2}}
\label{eq:12} 
\end{equation}

Note that   combining significances for experiments with different cuts in general does not help in search for new physics.
Really suppose we look for the number of isolated muons with transverse momentum:

(a). $  100~GeV < p_T  < 200~GeV$,

(b). $p_T >  200~GeV $,

(c). $p_T > 100~GeV$.

It is evident that the combination of the (a) ``experiment'' with the (b) ``experiment'' is 
equivalent to the (c) ``experiment''. And it is not clear which experiment (a), (b) or (c) will 
give the biggest significance (the biggest evidence in favour of new physics). The details will depend on 
the tested model of new physics.

This work was  supported by RFBR grant $\cal N$ 10-02-00468.

\end{document}